\documentclass[prr, aps, superscriptaddress, reprint]{revtex4-2}
\usepackage{lmodern}

\usepackage[T1]{fontenc}
\usepackage[latin9]{inputenc}
\usepackage{color}
\usepackage{amsmath}
\usepackage{amssymb}
\usepackage{graphicx}

\makeatletter

\pdfpageheight\paperheight
\pdfpagewidth\paperwidth

\makeatother

\usepackage{babel}
\begin{document}

\title{Single-Atom Amplification Assisted by Multiple Sideband Interference
in 1D Waveguide QED Systems}

\author{Kuan-Ting Lin}
	\thanks{These authors contribute equally.}
	\affiliation{Department of Physics and Center for Quantum Science and Engineering, National Taiwan University, Taipei 10617, Taiwan}
	\affiliation{Physics Division, National Center for Theoretical Sciences, Taipei 10617, Taiwan}

\author{Ting Hsu}
	\thanks{These authors contribute equally.}
	\affiliation{Department of Physics and Center for Quantum Science and Engineering, National Taiwan University, Taipei 10617, Taiwan}
	\affiliation{Physics Division, National Center for Theoretical Sciences, Taipei 10617, Taiwan}
	\affiliation{Trapped-Ion Quantum Computing Laboratory, Hon Hai Research Institute, Taipei 11492, Taiwan}

\author{Fahad Aziz}
	\affiliation{Department of Physics, National Tsing Hua University, Hsinchu 30013, Taiwan}

\author{Yu-Chen Lin}
	\affiliation{Department of Physics and Center for Quantum Science and Engineering, National Taiwan University, Taipei 10617, Taiwan}
	\affiliation{Physics Division, National Center for Theoretical Sciences, Taipei 10617, Taiwan}

\author{Ping-Yi Wen}
	\affiliation{Department of Physics, National Chung Cheng University, Chiayi
	621301, Taiwan}
\author{Io-Chun Hoi}
	\affiliation{Department of Physics, City University of Hong Kong, Hong Kong,
	China}

\author{Guin-Dar Lin}
	\email{guindarl@phys.ntu.edu.tw}
	\affiliation{Department of Physics and Center for Quantum Science and Engineering, National Taiwan University, Taipei 10617, Taiwan}
	\affiliation{Physics Division, National Center for Theoretical Sciences, Taipei 10617, Taiwan}
	\affiliation{Trapped-Ion Quantum Computing Laboratory, Hon Hai Research Institute, Taipei 11492, Taiwan}
	
\begin{abstract}
This study conducts a theoretical investigation into the signal amplification
arising from multiple Rabi sideband coherence within a one-dimensional
waveguide quantum electrodynamics system. We utilize a semi-infinite
waveguide to drive an anharmonic multi-level transmon with a strong
coherent microwave field, examining the scattering behavior by introducing
a probe signal. Our findings reveal signal amplification under specific
resonant conditions, presenting spectra that reveal finer details
than previously documented in the literature. To elucidate the mechanisms
behind this amplification, we develop a model that explicitly accounts
for multiple dressed sidebands in the presence of a strong driving
field. From this model, we derive the reflection amplitude of the
probe signal. Notably, our results indicate that amplification can
occur due to either population inversion or, in some instances, through
the constructive interference of multiple sidebands even in the absence
of population inversion. Additionally, we explore how qubit dephasing
impacts the amplification process.
\end{abstract}
\maketitle

\section{Introduction}

Quantum metrology relies on precise measurement of quantum systems
\citep{Giovannetti2011,Simon2017}, which is made possible through
the use of quantum amplifiers in the presence of a noisy environment
\citep{Colombo2022,Zuo2020,Hudelist2014,Xiang2010}. An effective
quantum amplifier requires strong coupling between its internal degrees
of freedom, such as atomic states, and the electromagnetic probe field.
It remains challenging with real atoms for their weak coupling to
the light, primarily caused by photonic spatial mode mismatch \citep{Wrigge2007,Tey2008,Leong2016,Gerhardt2007}.
This mismatch results in probabilistic absorption and low fidelity
in state manipulation, making the insertion of optical waveguides
and cavities a necessary step to boost the atom-photon interaction
\citep{Loo2013,Solano2017,Kien2005,Liao2015,Lin2019}. In the microwave
domain, the implementation of a one-dimensional (1D) waveguide system
involves the interaction of superconducting artificial atoms through
the guided modes of a 1D transmission line \citep{Sheremet2023}.
This system offers advantages in achieving strong coupling compared
to real atomic systems \citep{Gu2017}. Consequently, the 1D waveguide
system serves as an excellent platform for studying various quantum-optical
phenomena, including many-body entanglement \citep{Mirza2016,Mirza2016a,Facchi2016,Song2022,Zhang2023},
super-/sub-radiance, collective Lamb shift \citep{Mlynek2014,Lalumiere2013,Wen2019,Albrecht2019}
and applications related to quantum communications and information
processing \citep{Lin2022,Elfving2019,Mok2020,Liao2018}.

Furthermore, signal amplification has been recently investigated particularly
when a single atom is driven by a strong coherent field \citep{Wiegand2021,Chien2019,Wen2018,Koshino2013,Zhao2017,Vinu2020,Macklin2015}.
In this case, the energy levels of the bare atomic states mix with
the field to form dressed states, giving rise to spectral sidebands
known as Rabi sidebands \citep{Koshino2013,Mompart2000,Braumueller2015,Oelsner2013}.
While it is widely accepted that stimulated emissions contribute to
amplification of the probe signal when the probe field resonates with
the Rabi sidebands, previous studies have primarily focused on the
gain mechanisms strongly connected to population inversion associated
with a specific Rabi sideband. However, this perspective is valid
for weak and moderate pumping but overlook multiple sideband interference
when strong pumping is applied. Recently, the observation of signal
amplification originating from multiple Rabi sidebands coherence has
been reported \citep{Aziz2023}. Yet, the theoretical explanation
for this amplification and its mathematical formulation for strong-pumping
scenarios remains unclear.

In this work, we aim to theoretically investigate the detailed mechanisms
of gain in an amplifier composed of a transmon, which possesses anharmonic
multi-levels and is driven by a strong coherent microwave field. We
develop a theory that accounts for multiple Rabi sidebands and analyze
the effects associated with multiple photon processes. This general
approach not only applies to weak-pumping cases, where population
inversion plays an essential role in the gain, but also predicts signal
amplification resulting from quantum interference of multiple Rabi
sidebands in the presence of strong pumping. To address more realistic
conditions, we extend our discussion to include dephasing effects.
Furthermore, our framework can be generalized to include multiple
atoms, for which resonant dipole-dipole interactions come into play,
laying the groundwork for the conceptualization of a quantum many-body
amplifier.

This paper is organized as follows. In Sec.~\ref{sec: Theory}, we
present the theoretical model that considers a single transmon interacting
with a 1D semi-infinite waveguide, where we derive a multiple sideband
theory based on the dressed state representation. In Sec. ~\ref{sec: results},
we examine signal amplification in single-transmon cases due to one,
two, and three driving-photon processes of the driving field by analyzing
the reflection amplitude of the probe signal. The effects of pure
dephasing on the gain behavior are also discussed. Section~\ref{sec: Conclusion}
then connects this work with \citep{Aziz2023}, and concludes our
major findings.\textcolor{red}{{} }For completeness, we provide our
derivation details for more general multi-atom cases in the Appendix;
however, the associated discussion is beyond the scope of this work
and will be addressed elsewhere.

\section{Theory\label{sec: Theory}}

\subsection{Model and Hamiltonian}

\begin{figure}
\begin{centering}
\includegraphics[width=8.5cm]{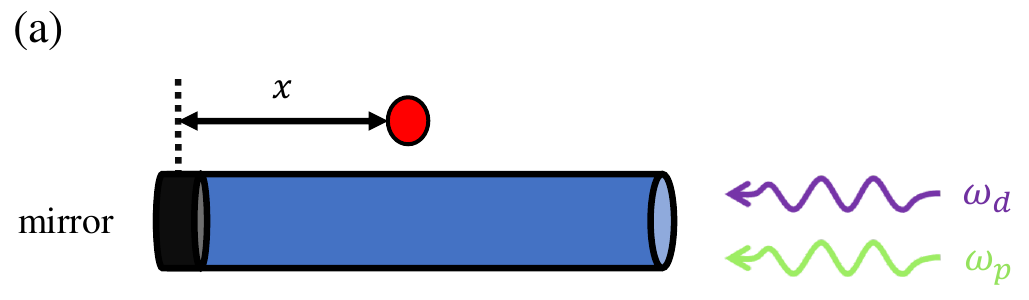}
\par\end{centering}
\begin{centering}
\includegraphics[width=8.5cm]{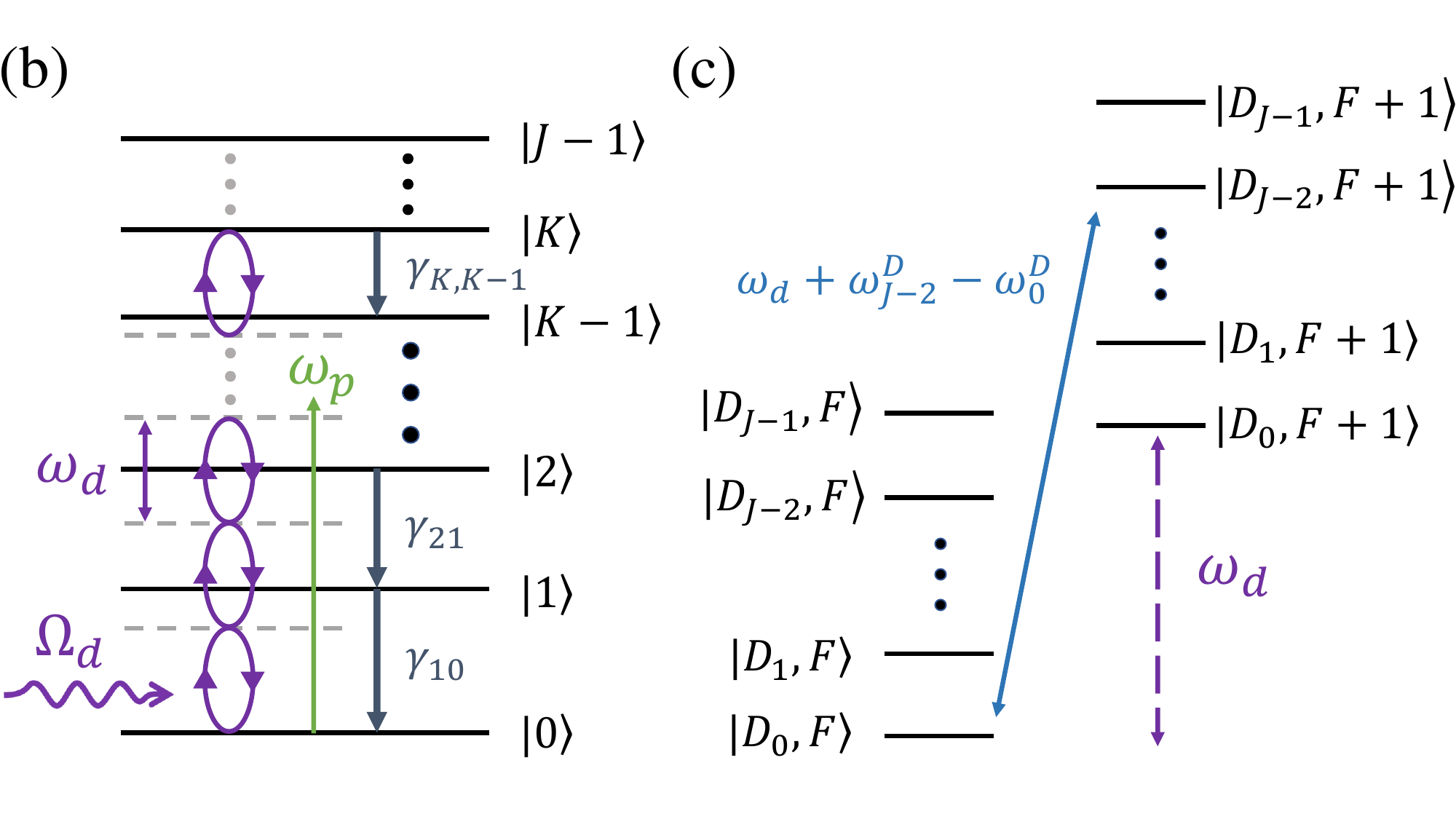}
\par\end{centering}
\caption{(a) The schematic illustrates an anharmonic multi-level transmon,
represented by a red circle, coupled to a semi-infinite 1D waveguide.
The terminated end of the waveguide acts as an antinode mirror located
at $x=0$. The coherent driving and probe fields are introduced from
the open end to interact with the transmon. (b) The energy diagram
representing a transmon with $J$ bare anharmonic levels along with
a strong coherent field characterized by the carrier and Rabi frequency
$\omega_{d}$ and $\Omega_{d}$, respectively. If $\omega_{d}$ is
chosen properly, a transmon can be pumped to the state $\left|K\right\rangle $
by absorbing $K$ photons. The system is then probed by a probe field
at frequency $\omega_{p}$. Here, the decay rate between adjacent
states $\left|j\right\rangle $ and $\left|j-1\right\rangle $ is
denoted by $\gamma_{j,j-1}$. (c) The dressed-state perspective: The
strong driving field mixes the bare atomic levels, yielding the dressed
states denoted by $\left|D_{\mu},F\right\rangle $ with $\mu=0,\cdots,J-1$
and $F$ associated with the photon number of the driving field.\label{fig: Model}}
\end{figure}

We consider a transmon artificial atom coupled to a 1D semi-infinite
waveguide with one end terminated by a perfect mirror \citep{Hoi2015},
as depicted in Fig.~\ref{fig: Model}(a). A transmon possesses multiple
anharmonic energy levels and we hereby impose a cutoff by considering
only the lowest $J$ levels, corresponding to energy $\hbar\omega_{j}$
for $j=1,2,\ldots,J$. We send in two-tone coherent fields, characterized
by the Rabi frequencies $\Omega_{d}$ and $\Omega_{p}$, and carrier
frequencies $\omega_{d}$ and $\omega_{p}$, to drive and probe the
transmon array, respectively. Note that we must choose $J$ large
enough such that the population in the $J$th level is negligible
all the times for a given finite strength of the driving field so
that the cutoff can be justified.

To describe the spectroscopic behavior, we consider the system's overall
Hamiltonian: $H=H_{s}+H_{b}+V+H_{d}+H_{p}$ \citep{Peropadre2013,Koshino2012,Dorner2002},
where $H_{s}=\sum_{j}\hbar\omega_{j}\sigma_{jj}$ and $H_{b}=\hbar\int_{0}^{\infty}\omega a_{\omega}^{\dagger}a_{\omega}d\omega$
represent the transmons' and field's parts, respectively, with $\sigma_{jj}=\left|j\right\rangle \left\langle j\right|$
corresponding to the projection operator of the $j$th energy level
of the transmon; $a_{\omega}^{\dagger}$ and $a_{\omega}$ are the
creation and annihilation operators, respectively, of a waveguide
photon of frequency $\omega$. Note that the commutation relation
$\left[a_{\omega},a_{\omega^{\prime}}^{\dagger}\right]=\delta\left(\omega-\omega^{\prime}\right)$
is satisfied.

The interaction between the transmon and the waveguide modes is described
by
\begin{equation}
V=i\hbar\sum_{j}\int_{0}^{\infty}d\omega\sqrt{j}g\left(\omega\right)\cos\left(k_{\omega}x\right)\sigma_{j,j-1}a_{\omega}+{\rm H.c.}\label{eq: interaction H}
\end{equation}
In this expression, $\sigma_{j,k}=\left|j\right\rangle \left\langle k\right|$
represents the transition operator between the $j$th and $k$th levels
of the transmon located at position $x$. The coupling strength $\sqrt{j}g\left(\omega\right)$
describes the interaction between a transmon around the $j$th level
and a waveguide photon of frequency $\omega$ \citep{Koch2007}. The
phase seen by the transmon is modulated by $\cos\left(k_{\omega}x\right)$
given the boundary condition of an antinode at $x=0$ \citep{Wen2019,Lin2019},
where $k_{\omega}=\omega/v_{g}$ represents the wavenumber with the
speed of light $v_{g}$; ${\rm H.c.}$ stands for the Hermitian conjugate.
Lastly, the driving and probe fields coupling to this transmon are
described respectively by \citep{Kockum2013}, 
\begin{equation}
H_{d}=\hbar\sum_{j}\frac{\sqrt{j}\Omega_{d}\cos\left(k_{d}x\right)}{2}\left(\sigma_{j,j-1}e^{-i\omega_{d}t}+{\rm H.c.}\right)\label{eq: drvign H}
\end{equation}
and 
\begin{equation}
H_{p}=\hbar\sum_{j}\frac{\sqrt{j}\Omega_{p}\cos\left(k_{p}x\right)}{2}\left(\sigma_{j,j-1}e^{-i\omega_{p}t}+{\rm H.c.}\right).\label{eq: probe H}
\end{equation}
Here, $k_{d}=\omega_{d}/v_{g}$ and $k_{p}=\omega_{p}/v_{g}$ represent
the wavenumbers for the driving and probe fields, respectively. Following
the standard approach of tracing out the waveguide modes \citep{Carmichael1999,Lin2019},
we then obtain the Born-Markov master equation, which, in the rotating
frame of multiple frequencies of $\omega_{d}$, reads
\begin{equation}
\begin{aligned}\frac{d\rho}{dt} & =i\sum_{j}\left(j\omega_{d}-\omega_{j}\right)\left[\sigma_{jj},\rho\right]\\
 & +i\sum_{j}\frac{\sqrt{j}\Omega_{d}\cos\left(k_{d}x\right)}{2}\left(\left[\sigma_{j,j-1},\rho\right]-{\rm H.c.}\right)\\
 & +i\sum_{j}\frac{\sqrt{j}\Omega_{p}\cos\left(k_{p}x\right)}{2}\left(\left[\sigma_{j,j-1},\rho\right]e^{i\left(\omega_{d}-\omega_{p}\right)t}-{\rm H.c.}\right)\\
 & +\sum_{jl}\frac{\sqrt{jl}\gamma_{j,j-1}}{2}\left(\left[\sigma_{j-1,j}\rho,\sigma_{l,l-1}\right]+{\rm H.c.}\right)\\
 & +\sum_{j}\gamma_{j}^{\phi}\left(\left[\sigma_{jj}\rho,\sigma_{jj}\right]+{\rm H.c.}\right).
\end{aligned}
\label{eq: master eq}
\end{equation}
Here, $\rho$ is the reduced $J\times J$ density matrix dealing merely
with the transmon's atomic degrees of freedom. The relaxation of the
atomic level is characterized by $\gamma_{j,j-1}=\gamma_{j,j-1}^{(0)}\cos^{2}k_{j,j-1}x$,
where $\gamma_{j,j-1}^{(0)}=2\pi g^{2}\left(\omega_{j,j-1}\right)$
represents the bare decay rate at the transition frequency $\omega_{j,j-1}=\omega_{j}-\omega_{j-1}$,
and $k_{j,j-1}=\omega_{j,j-1}/v_{g}$ is the associated wavenumber.
The last line of the master equation (\ref{eq: master eq}) is inserted
to account for the pure dephasing effect, characterized by the dephasing
rate $\gamma_{j}^{\phi}$ of the $j$th level. It is noteworthy that
the main interest of this work is to investigate the probe amplification
near the $j$-photon resonance, or more specifically, $j\omega_{d}-\omega_{j}\ll\omega_{j,j-1}$
{[}see Fig.~\ref{fig: Model}(b){]}, which validates the rotating-wave
approximation in the master equation (\ref{eq: master eq}).

\subsection{Reflection amplitude: Perturbative analysis}

To study the reflection of the probe field, we define the reflection
amplitude $r=\left|\left\langle a_{{\rm out}}\left(t\right)\right\rangle /\left\langle a_{{\rm in}}\left(t\right)\right\rangle \right|$
with the photonic operators $a_{{\rm out}}\left(t\right)$ and $a_{{\rm in}}\left(t\right)$
representing the output and input, respectively, of the signal, i.e.,
the probe field. The output signal can be determined through the input-output
relation \citep{Gardiner1985,Lalumiere2013}, 
\begin{equation}
a_{{\rm out}}\left(t\right)=a_{{\rm in}}\left(t\right)+\sum_{j}\sqrt{j\gamma_{j,j-1}}\cos\left(k_{j,j-1}x\right)\sigma_{j-1,j}\left(t\right).\label{eq: input_output relation}
\end{equation}
In our setup, we send a single-mode classical probe field as the input
signal so that the operator $a_{{\rm in}}\left(t\right)$ can be replaced
by \citep{Strandberg2019,Lu2021}
\begin{equation}
a_{{\rm in}}\rightarrow\frac{i\Omega_{p}}{2\sqrt{\omega_{p}\gamma_{10}/\omega_{10}}}e^{-i\omega_{p}t}.\label{eq: classical input}
\end{equation}
Combining Eqs.~(\ref{eq: input_output relation}) and (\ref{eq: classical input})
leads to

\begin{equation}
r=\left|1+\frac{2i\sum_{j}\tilde{\gamma}_{j}\left(x\right)\left\langle \tilde{\sigma}_{j-1,j}\left(t\right)\right\rangle e^{i\left(\omega_{p}-\omega_{d}\right)t}}{\Omega_{p}}\right|,\label{eq: reflection}
\end{equation}
where $\tilde{\sigma}_{j-1,j}\left(t\right)=\sigma_{j-1,j}\left(t\right)e^{i\omega_{d}t}$
representing the corresponding ladder operator in the rotating frame
of frequency $\omega_{d}$, and $\tilde{\gamma}_{j}\left(x\right)=\sqrt{j\omega_{p}\gamma_{10}\gamma_{j,j-1}/\omega_{10}}\cos\left(k_{j,j-1}x\right)$. 

Usually, the probe field is taken to be sufficiently weak, i.e. $\Omega_{p}/\gamma_{10},\Omega_{p}/\Omega_{d}\ll1$.
One can thus approximate 

\begin{equation}
\rho\left(t\right)\approx\rho^{\left(0\right)}\left(t\right)+\frac{\Omega_{p}}{\gamma_{10}}\rho^{\left(1\right)}\left(t\right)e^{-i\left(\omega_{p}-\omega_{d}\right)t},\label{eq: expansion rho}
\end{equation}
where $\rho^{(0)}$ is the $0$th-order solution in the absence of
the probe field, and $\rho^{(1)}$ accounts for the leading (first)
order contribution proportional to $\Omega_{p}$. We then solve for
the dynamics of the density operators $\rho^{\left(0\right)}$ and
$\rho^{\left(1\right)}$ by substituting Eq.~(\ref{eq: expansion rho})
into Eq.~(\ref{eq: master eq}), and determine the expectation value
$\left\langle \tilde{\sigma}_{j-1,j}\left(t\right)\right\rangle ={\rm Tr}\left(\Omega_{p}\tilde{\sigma}_{j-1,j}\rho^{\left(1\right)}\left(t\right)e^{-i\left(\omega_{p}-\omega_{d}\right)t}/\gamma_{10}\right)$.

\subsection{Dressed-state approach: Multi-sideband model }

In this section, we present the dressed-state perspective for strong-driving
scenarios, where $H_{s}+H_{d}$ plays a dominant role and can be diagonalized
to form dressed states. Taking the associated photon excitation into
account, the dressed states can be represented by $\left|D_{\mu},F\right\rangle $
with the eigenenergy $\hbar\left(\omega_{\mu}^{D}+F\omega\right)$,
as depicted in Fig.~\ref{fig: Model}(c). Here, $\left|D_{\mu}\right\rangle $
is an eigenstate of the interaction-picture Hamiltonian of $H_{s}+H_{d}$
with $\mu=0,\cdots,J-1$ labeling the dressed states in the ascending
order of eigenenergies $\omega_{\mu}^{D}$ and $F$ denoting the photon
number of the driving field \citep{Koshino2013}. Note that these
dressed-state eigenenergies are viewed in their specific rotating
frames, their ordering does not reflect that according to the true
energies in the bare-state basis. For example, the bare ground state
$|0\rangle$ might be mapped to the dressed-state $|D_{\mu=J-1}\rangle$
when the pump field becomes intensive. For clarity, in this manuscript,
we use Greek letters for the sub-indices labeling the dressed states
such as $D_{\mu}$, and English letters for the sub-indices labeling
the bare states.

The reflection amplitude can then be written as
\begin{equation}
r=\left|1+2i\sum_{\mu\nu}C_{D_{\mu}D_{\nu}}\left\langle \sigma_{D_{\mu}D_{\nu}}\left(t\right)\right\rangle \right|.\label{eq: dresed stated reflection}
\end{equation}
The coefficient $C_{D_{\mu}D_{\nu}}$ is related to the atomic ladder
operator through 
\begin{equation}
\begin{aligned}C_{D_{\mu}D_{\nu}} & =\sum_{j}\frac{\tilde{\gamma}_{j}\left(x\right)}{\gamma_{10}}{\rm Tr}\left(\sigma_{j-1,j}\sigma_{D_{\nu}D_{\mu}}\right),\end{aligned}
\label{eq:coeifficient Cuv}
\end{equation}
where $\sigma_{D_{\nu}D_{\mu}}=\left|D_{\nu}\right\rangle \left\langle D_{\mu}\right|=\sigma_{D_{\mu}D_{\nu}}^{\dagger}$,
and its dynamics are determined through the optical Bloch equations:\begin{widetext}
\begin{equation}
\begin{aligned}\frac{d\left\langle \sigma_{D_{\mu}D_{\nu}}\right\rangle }{dt}=\frac{d\left\langle D_{\nu}\right|\rho^{\left(1\right)}\left|D_{\mu}\right\rangle }{dt} & =\left(i\delta_{\mu\nu}^{D}+\Gamma_{\mu\nu}^{D}\right)\left\langle \sigma_{D_{\mu}D_{\nu}}\right\rangle \\
 & +i\sum_{\eta}\left(\Omega_{\eta\nu}^{D}\left\langle \sigma_{D_{\mu}D_{\eta}}\right\rangle _{\left(0\right)}-\Omega_{\mu\eta}^{D}\left\langle \sigma_{D_{\eta}D_{\nu}}\right\rangle _{\left(0\right)}\right)\\
 & +\sum_{\xi\neq\mu,\eta\neq\nu}\bar{\Gamma}_{\eta\nu\mu\xi}^{D}\left\langle \sigma_{D_{\xi}D_{\eta}}\right\rangle -\sum_{\eta,\xi\neq\nu}\Gamma_{\eta\nu\xi\eta}^{D}\left\langle \sigma_{D_{\mu}D_{\xi}}\right\rangle -\sum_{\xi,\eta\neq\mu}\Gamma_{\xi\eta\mu\xi}^{D}\left\langle \sigma_{D_{\eta}D_{\nu}}\right\rangle 
\end{aligned}
\label{eq:OBE}
\end{equation}
\end{widetext} with $\delta_{\mu\nu}^{D}=\omega_{p}-\left(\omega_{\nu}^{D}-\omega_{\mu}^{D}+\omega_{d}\right)$
denoting the relevant detuning in the dressed-state representation.
The relaxation rate is expressed as $\Gamma_{\mu\nu}^{D}=\bar{\Gamma}_{\nu\nu\mu\mu}^{D}-\sum_{\eta}\left(\Gamma_{\eta\nu\nu\eta}^{D}+\Gamma_{\eta\mu\mu\eta}^{D}\right)$
with
\begin{equation}
\begin{aligned}\bar{\Gamma}_{\eta\nu\mu\xi}^{D} & =\sum_{jk}\chi_{jk}{\rm Tr}\left[\sigma_{k-1,k}\sigma_{D_{\eta}D_{\nu}}\right]{\rm Tr}\left[\sigma_{j,j-1}\sigma_{D_{\mu}D_{\xi}}\right]\\
 & \qquad+2\sum_{j}\gamma_{j}^{\phi}{\rm Tr}\left[\sigma_{jj}\sigma_{D_{\eta}D_{\nu}}\right]{\rm Tr}\left[\sigma_{jj}\sigma_{D_{\mu}D_{\xi}}\right]
\end{aligned}
\label{eq:Gamma_bar}
\end{equation}
and 
\begin{equation}
\begin{aligned}\Gamma_{\eta\nu\mu d}^{D} & =\sum_{jk}\eta_{jk}{\rm Tr}\left[\sigma_{j,j-1}\sigma_{D_{\eta}D_{\nu}}\right]{\rm Tr}\left[\sigma_{k-1,k}\sigma_{D_{\mu}D_{\xi}}\right]\\
 & +\sum_{j}\gamma_{j}^{\phi}{\rm Tr}\left[\sigma_{jj}\sigma_{D_{\eta}D_{\nu}}\right]{\rm Tr}\left[\sigma_{jj}\sigma_{D_{\mu}D_{\xi}}\right]
\end{aligned}
\label{eq:Gamma_pn}
\end{equation}
where $\chi_{jk}=\sqrt{jk}\left(\frac{\gamma_{k,k-1}+\gamma_{j,j-1}}{2}\right)$
and $\eta_{jk}=\sqrt{jk}\left(\frac{\gamma_{k,k-1}}{2}\right)$. Note
that the second row of the right-hand side of Eq.~(\ref{eq:OBE})
describes the pumping process with $\Omega_{\mu\nu}^{D}\left\langle \sigma_{D_{\mu}D_{\nu}}\right\rangle _{\left(0\right)}=\sum_{j}\frac{\sqrt{j}\gamma_{10}}{2}\cos\left(k_{p}x\right){\rm Tr}\left[\sigma_{j,j-1}\sigma_{D_{\mu}D_{\nu}}\right]{\rm Tr}\left[\rho^{\left(0\right)}\left|D_{\nu}\right\rangle \left\langle D_{\mu}\right|\right]$
while the third row corresponds to the coupling between the dressed
states.

\subsection{Resonant responses \label{subsec:Resonant-responses}}

The approach discussed in the preceding section describes how we calculate
the reflection amplitude in general cases. As we will see the spectral
results in Sec.~\ref{sec: results}, one can in fact identify the
correspondence between the spectral curves to certain dressed-state
transitions. To understand such correspondence, we expect that only
the near-resonant sidebands contribute significantly to the amplification
or attenuation of the reflection signal. Here, by near-resonant we
mean that the condition with detuning $\delta_{\mu\nu}^{D}\lesssim\gamma_{10}$
is fulfilled for $\mu,\nu=0,1,\ldots,J-1$ with $\mu\neq\nu$.

The associated mechanisms can be described by a reduced model in which
we only keep the effective transitions, which are supposed to have
interfering effects on the reflection spectrum \citep{Aziz2023}.
For example, considering two such Rabi sideband transitions, say $\left|D_{\nu_{1}},F+1\right\rangle \leftrightarrow\left|D_{\mu_{1}},F\right\rangle $
and $\left|D_{\nu_{2}},F+1\right\rangle \leftrightarrow\left|D_{\mu_{2}},F\right\rangle $,
resonant with the probe signal, from Eq.~(\ref{eq:OBE}) we obtain
the following coupled equations for the steady state:\begin{widetext}
\begin{equation}
\left[\begin{array}{cc}
\Gamma_{\mu_{1}\nu_{1}}^{D}+i\delta_{\mu_{1}\nu_{1}}^{D} & \Gamma_{\nu_{2}\nu_{1}\mu_{1}\mu_{2}}^{D}\\
\Gamma_{\nu_{1}\nu_{2}\mu_{2}\mu_{1}}^{D} & \Gamma_{\mu_{2}\nu_{2}}^{D}+i\delta_{\mu_{2}\nu_{2}}^{D}
\end{array}\right]\left[\begin{array}{c}
\left\langle \sigma_{D_{\mu_{1}}D_{\nu_{1}}}\right\rangle \\
\left\langle \sigma_{D_{\mu_{2}}D_{\nu_{2}}}\right\rangle 
\end{array}\right]=\left[\begin{array}{c}
i\sum_{\eta}\left(\Omega_{\mu_{1}\eta}^{D}\left\langle \sigma_{D_{\eta}D_{\nu_{1}}}\right\rangle _{\left(0\right)}-\Omega_{\eta\nu_{1}}^{D}\left\langle \sigma_{D_{\mu_{1}}D_{\eta}}\right\rangle _{\left(0\right)}\right)\\
i\sum_{\eta}\left(\Omega_{\mu_{2}\eta}^{D}\left\langle \sigma_{D_{\eta}D_{\nu_{2}}}\right\rangle _{\left(0\right)}-\Omega_{\eta\nu_{2}}^{D}\left\langle \sigma_{D_{\mu_{2}}D_{\eta}}\right\rangle _{\left(0\right)}\right)
\end{array}\right].\label{eq: double_sidebands_EQ}
\end{equation}
\end{widetext} The non-diagonal terms in the matrix describe the
interference between the two Rabi sideband transitions, establishing
correlations among the dressed states. It becomes crucial to account
for such interference in order to calculate the actual reflection
amplitude especially when the pumping is intensive.

\section{results\label{sec: results} }

\subsection{Single-photon resonance and amplification\label{subsec:One-photon-process}}

\begin{figure}
\begin{centering}
\includegraphics[width=6cm]{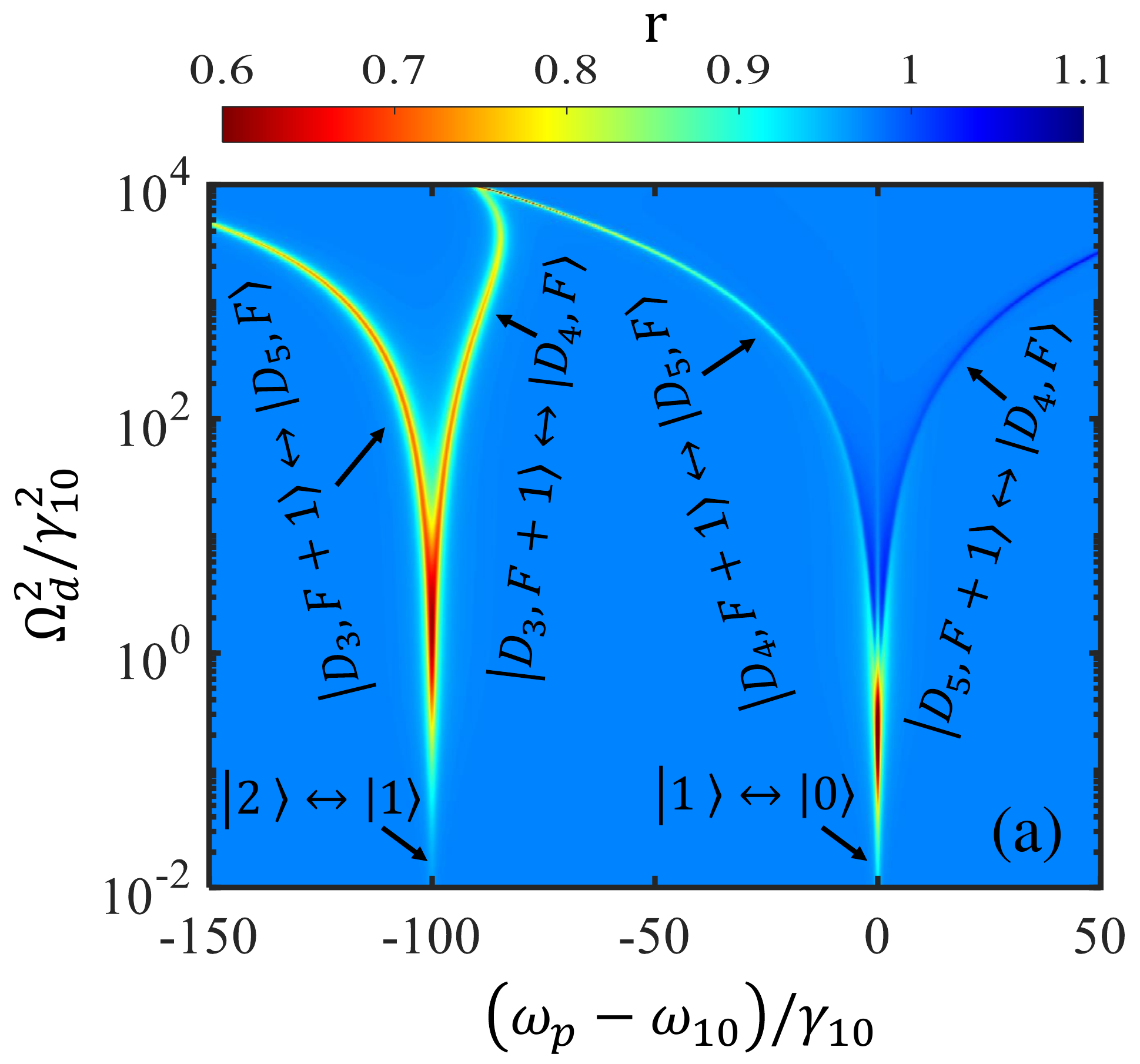}
\par\end{centering}
\begin{centering}
\includegraphics[width=6cm]{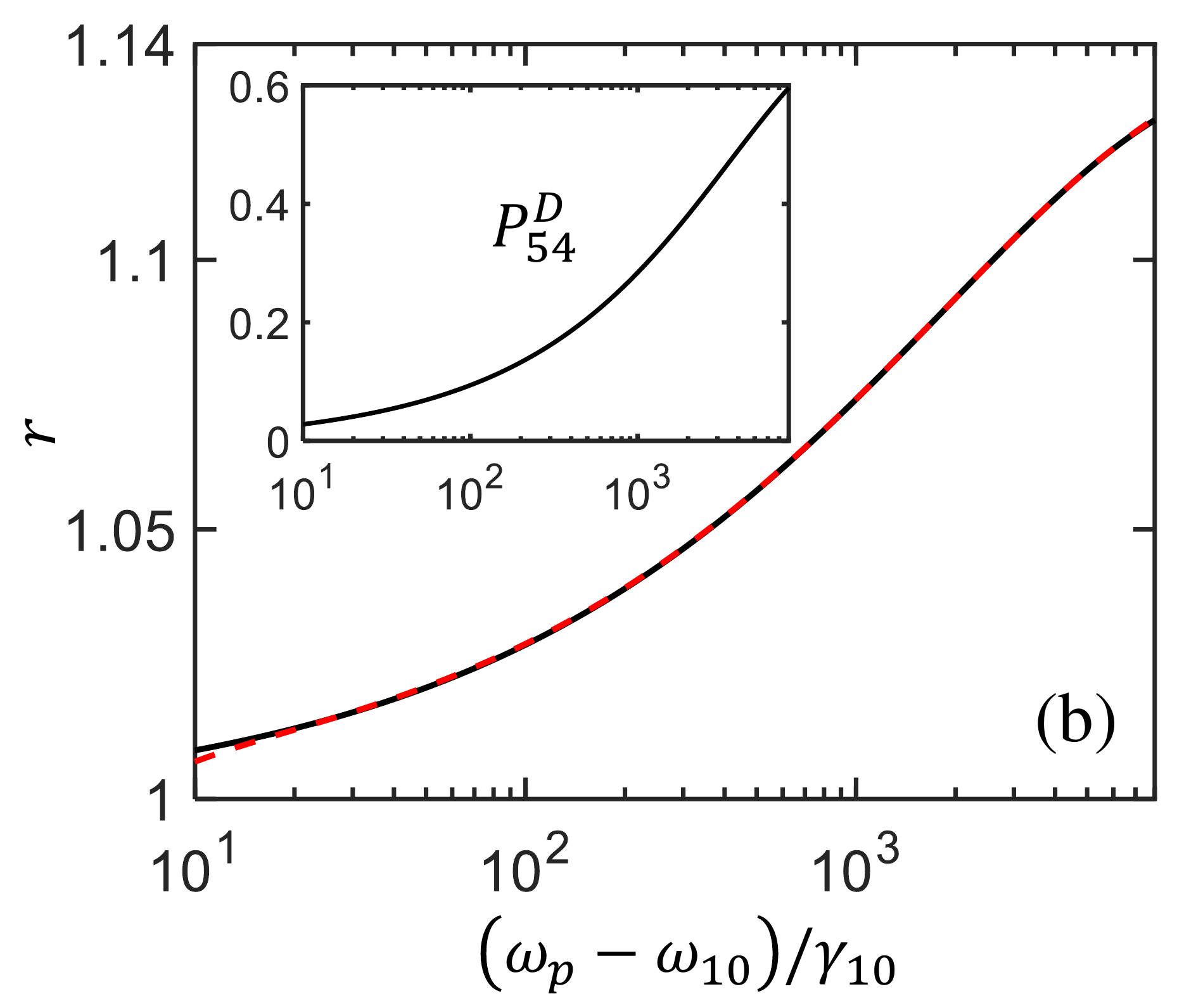}
\par\end{centering}
\caption{(a) Reflection amplitude for varied driving powers $\Omega_{d}^{2}/\gamma_{10}^{2}$
and probe frequencies $\omega_{p}$. In this simulation, we set $J=6$
and use the decay rates $\gamma_{n,n-1}=n\gamma_{10}$ and the transmon
energy $\hbar\omega_{n}=n\hbar\omega_{10}-\frac{n\left(n-1\right)}{2}\hbar\alpha$,
where the frequency $\omega_{10}=2100\gamma_{10}$, the anharmonicity
$\alpha=100\gamma_{10}$ and $n=1,2,\ldots,5$. (b) The reflection
amplitude corresponding to the $\left|D_{5},F+1\right\rangle \leftrightarrow\left|D_{4},F\right\rangle $
transition. The amplitude values computed by our reduced resonant
model and full master equation are represented by the red and black
curves, respectively. The inset illustrates the population difference
$P_{54}^{D}=\left\langle \sigma_{D_{5}D_{5}}\right\rangle _{\left(0\right)}-\left\langle \sigma_{D_{4}D_{4}}\right\rangle _{\left(0\right)}$.
\label{fig: single-sideband}}
\end{figure}

We first look at the simplest nontrivial case by setting the driving
frequency $\omega_{d}=\omega_{10}=\omega_{1}-\omega_{0}$, displaying
the single pump photon effect. This system is then probed, and the
associated reflection amplitude is shown in Figure~\ref{fig: single-sideband}(a).
Here, we find that the results for $J\ge6$ coincide with negligible
discrepancy so that we choose the cut-off $J=6$. When the driving
strength $\Omega_{d}^{2}/\gamma_{10}^{2}\lesssim10^{0}$ is considered
weak, we observe one bright stripe associated with the $\left|1\right\rangle \leftrightarrow\left|0\right\rangle $
transition and another with the $\left|2\right\rangle \leftrightarrow\left|1\right\rangle $
transition at frequencies $\omega_{p}=\omega_{10}$ and $\omega_{p}-\omega_{10}=-\alpha=-100\gamma_{10}$,
respectively, where $\alpha\equiv\omega_{10}-\omega_{21}$ is the
level anharmonicity of the transmon. 

As the driving strength increases and enters the regime $\Omega_{d}^{2}/\gamma_{10}^{2}\gtrsim10^{0}$,
the $\left|1\right\rangle \leftrightarrow\left|0\right\rangle $ transition
starts to split into two branches, signaling the formation of the
dressed states. Specifically, the right-branch corresponds to the
transition $|D_{5},F+1\rangle\leftrightarrow|D_{4},F\rangle$ while
the left-branch to the transition $|D_{4},F+1\rangle\leftrightarrow|D_{5},F\rangle$.
Similarly, when $\Omega_{d}^{2}/\gamma_{10}^{2}\gtrsim10^{1}$, the
$\left|2\right\rangle \leftrightarrow\left|1\right\rangle $ transition
splits because the state $|2\rangle$ starts to be populated, with
the right-branch corresponding to $|D_{3},F+1\rangle\leftrightarrow|D_{4},F\rangle$
and the left-branch to $|D_{3},F+1\rangle\leftrightarrow|D_{5},F\rangle$.
Further, we observe both attenuation ($r<1$) and amplification ($r>1$)
of the probe field along these resonant paths. We want to highlight
that the transition $|D_{5},F+1\rangle\leftrightarrow|D_{4},F\rangle$
presenting amplification is directly related to population inversion,
consistent with \citep{Koshino2013}.

In order to elucidate the gain mechanism given that the driving power
is high enough ($\Omega_{d}^{2}/\gamma_{10}^{2}\gtrsim10^{1}$), we
can determine the steady-state value of $\left\langle \sigma_{D_{\mu}D_{\nu}}\right\rangle $
using Eq.~(\ref{eq:OBE}), and obtain the reflection amplitude of
the probe signal: 
\begin{equation}
r=\left|1-\frac{2C_{D_{\mu}D_{\nu}}\Omega_{\mu\nu}^{D}P_{\nu\mu}^{D}}{\Gamma_{\mu\nu}^{D}}\right|\label{eq: single-sidebnad-general}
\end{equation}
with $P_{\nu\mu}^{D}=\left\langle \sigma_{D_{\nu}D_{\nu}}\right\rangle _{\left(0\right)}-\left\langle \sigma_{D_{\mu}D_{\mu}}\right\rangle _{\left(0\right)}$
denoting the population difference between the two states $\left|D_{\nu}\right\rangle $
and $\left|D_{\mu}\right\rangle $. It can be verified that the factor
$2C_{D_{\mu}D_{\nu}}\Omega_{\mu\nu}^{D}/\Gamma_{\mu\nu}^{D}$ is real
and negative for all parameter regimes of interest, suggesting the
direct implication that the amplification of the probe field ($r>1$)
is linked to the population inversion ($P_{\nu\mu}^{D}>1$) \citep{Koshino2013,Wiegand2021,Astafiev2010,Chien2019}. 

In Fig.~\ref{fig: single-sideband}(b), we compare the reflection
amplitudes computed by the full master equation Eq.~(\ref{eq: master eq})
and the reduced resonant model described in sub-Sec.~\ref{subsec:Resonant-responses}.
We find excellent agreement between the two approaches. The associated
population inversion in the relevant dressed states is also explicitly
shown. We can thus expect that the amplification in the probe signal
originates from the stimulated emission of the Rabi sideband.

\subsection{Multi-photon resonance and enhanced amplification \label{subsec:Multi-photon-process}}

\begin{figure*}
\begin{centering}
\includegraphics[width=5.5cm]{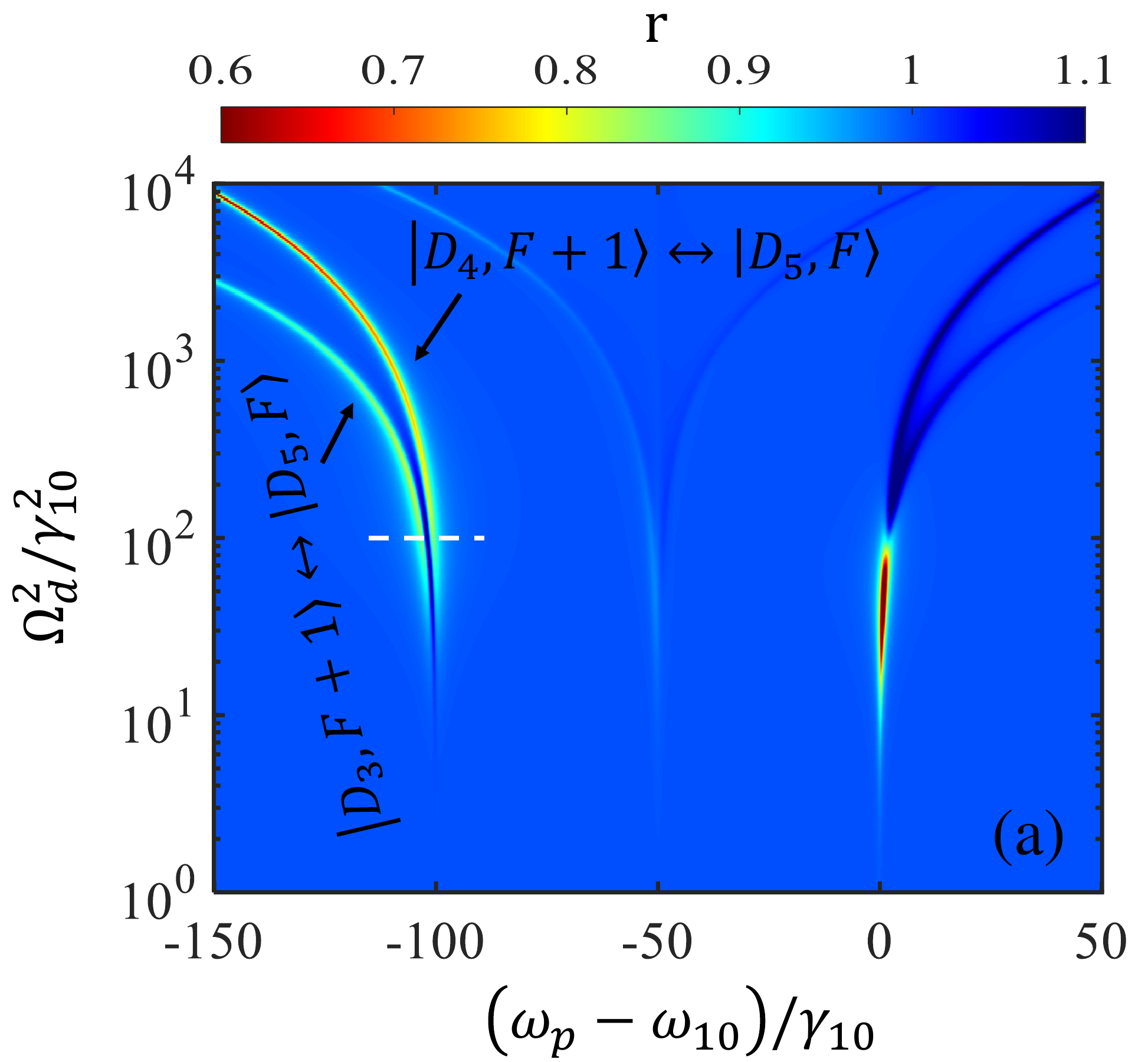}\includegraphics[width=5.5cm]{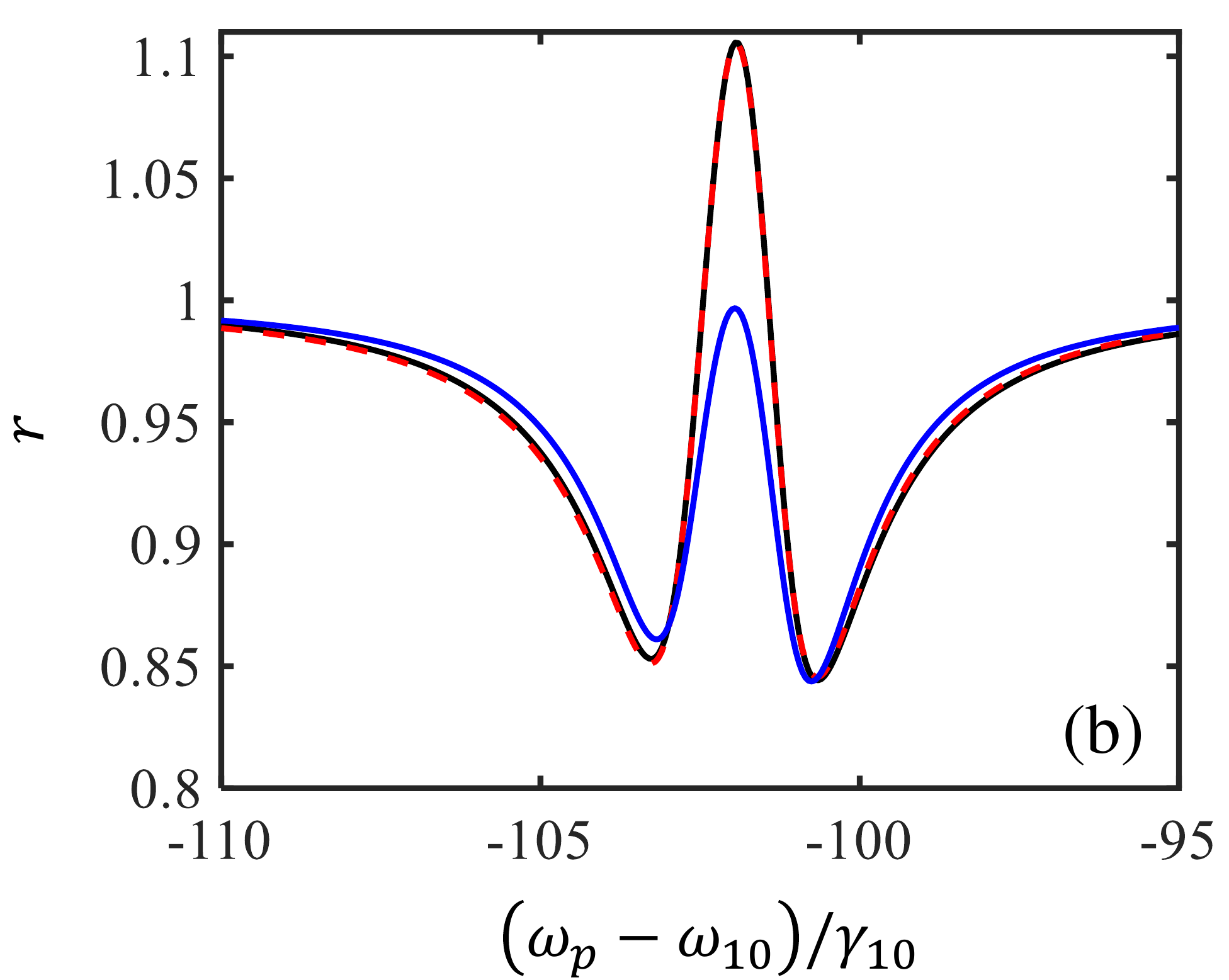}\includegraphics[width=5.5cm]{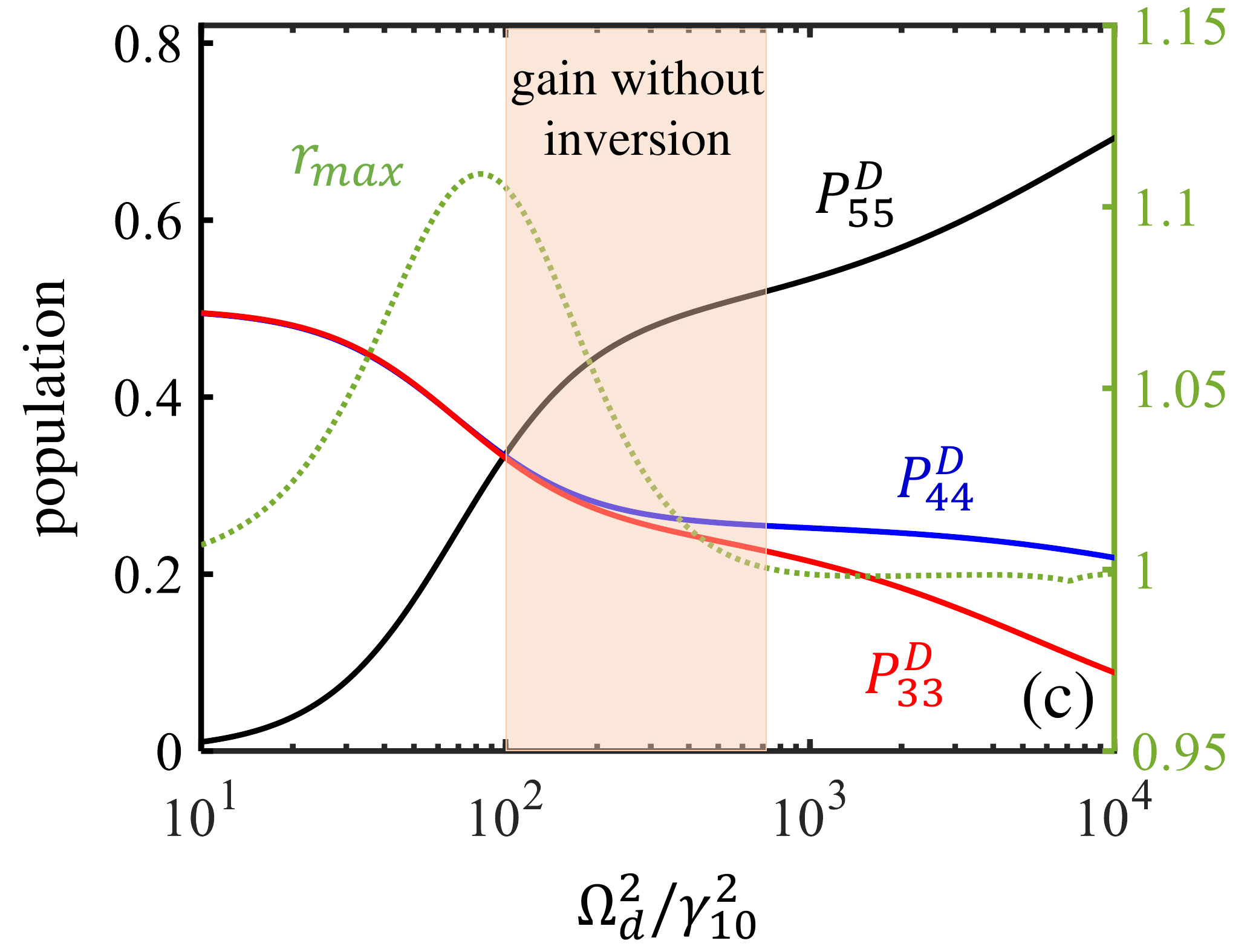}
\par\end{centering}
\begin{centering}
\includegraphics[width=5.5cm]{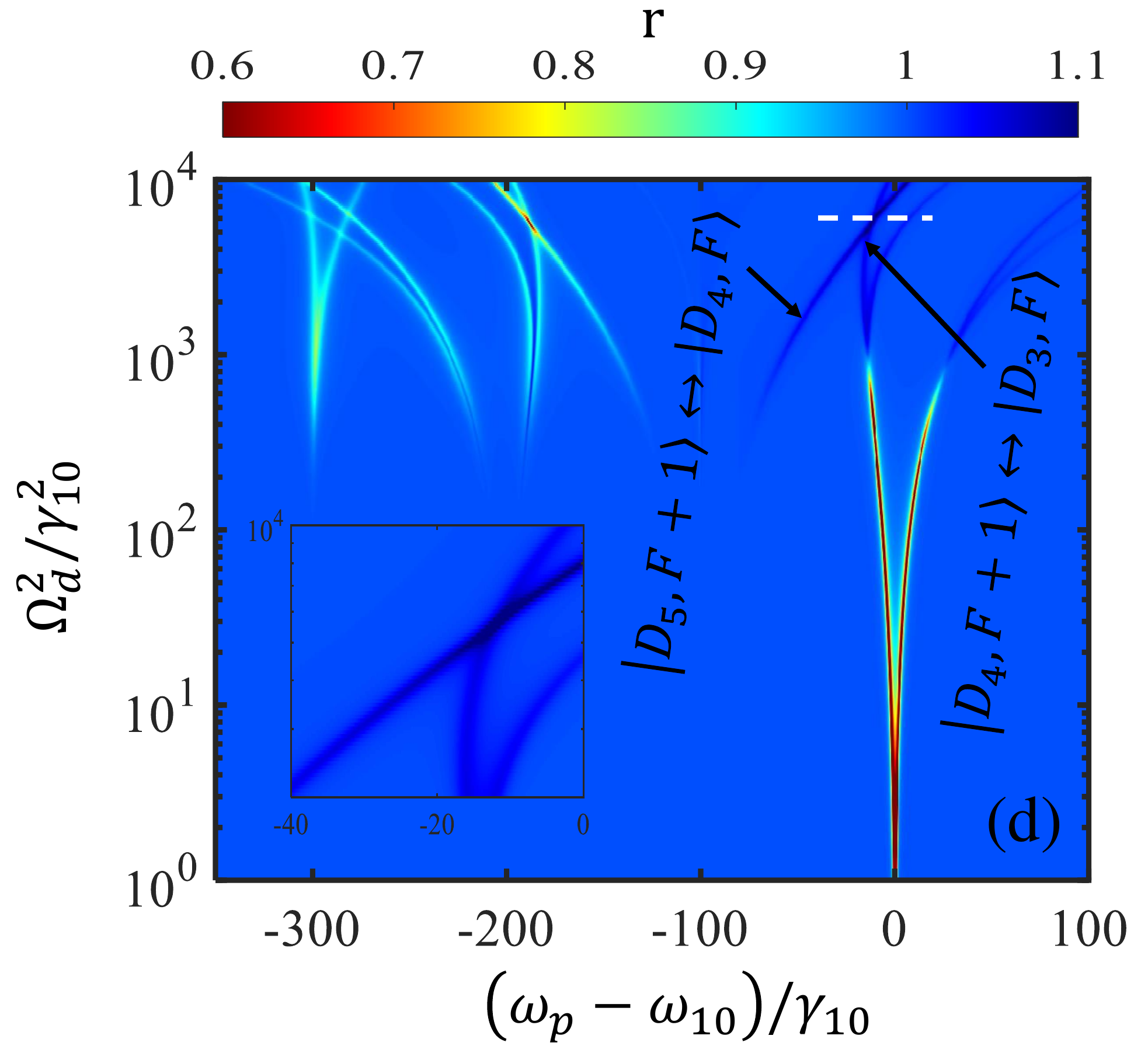}\includegraphics[width=5.5cm]{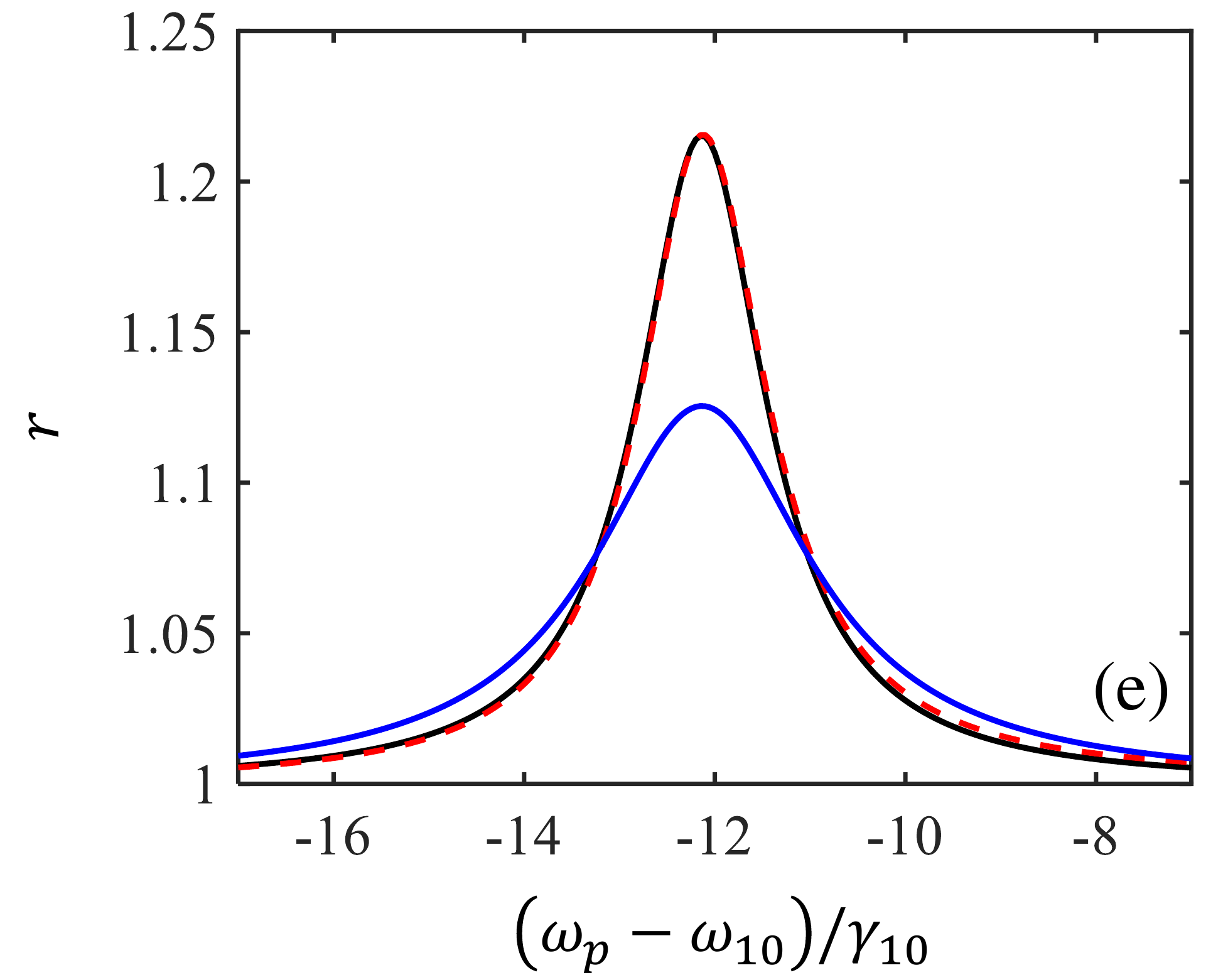}\includegraphics[width=5.5cm]{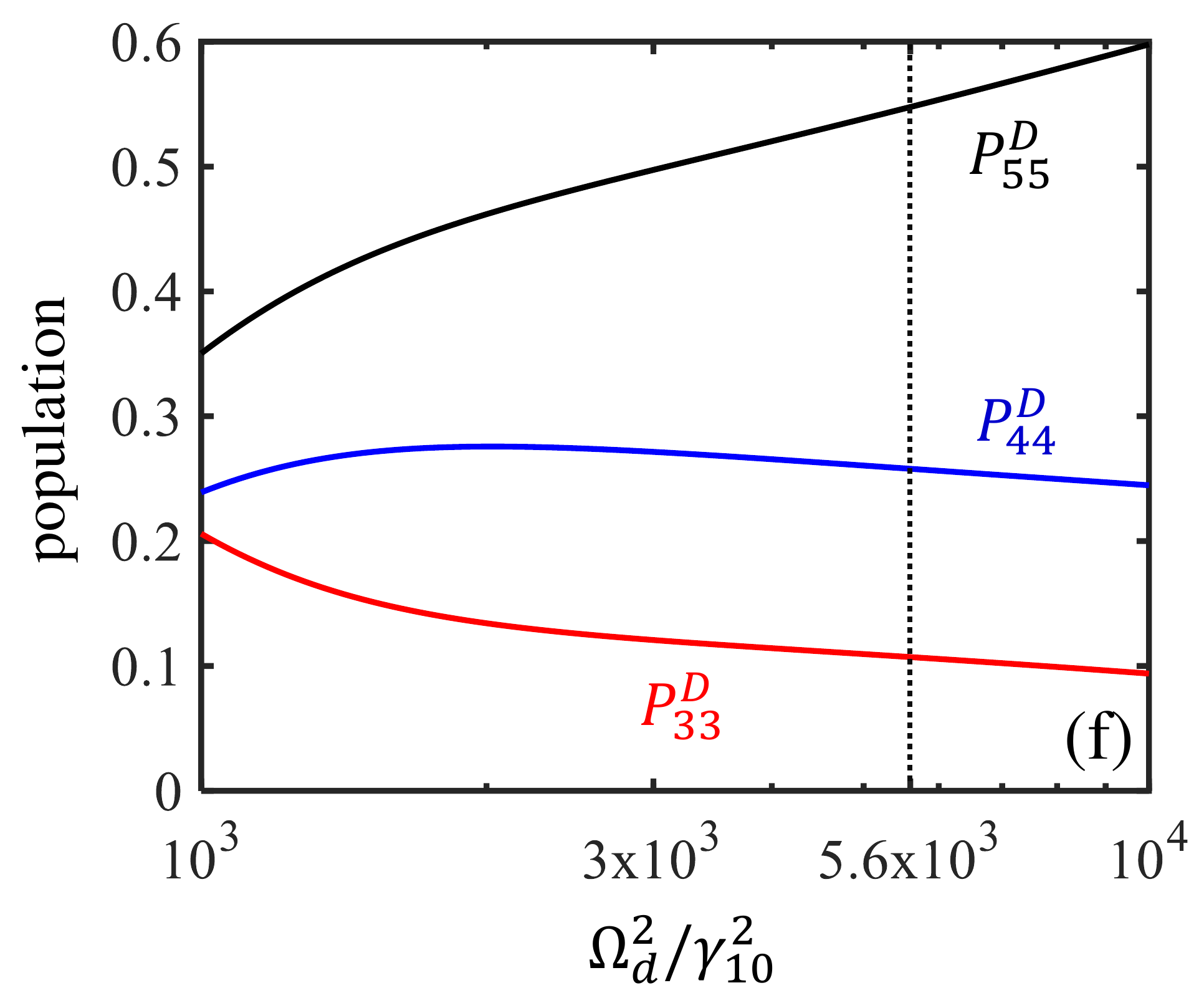}
\par\end{centering}
\caption{Reflection amplitudes and population distributions near $K$ pump-photon
resonances for (a)--(c) $K=2$ and (d)--(f) $K=3$. (a) and (d)
show the reflection amplitudes $r$ as a function of the driving power
$\Omega_{d}^{2}/\gamma_{10}^{2}$ and the probe frequency $\omega_{p}$,
obtained by the full master equation approach. We label the relevant
Rabi sideband transitions along with the curves where we find enhanced
amplifications. In the inset of (d) is the zoomed-in plot when two
sideband transitions cross each other. (b) and (e) show the reflection
profiles following the line cuts (white dashed lines) in (a) and (d),
where the driving powers are $\Omega_{d}^{2}/\gamma_{10}^{2}=10^{2}$
and $\Omega_{d}^{2}/\gamma_{10}^{2}=5.6\times10^{3}$, respectively.
For comparison, we plot all three reflection amplitude curves computed
by the full master equation (red dashed), the effective Rabi sideband
model Eq.~(\ref{eq: double_sidebands_EQ}) (black solid), and the
simplified version without the coupling between the dressed-state
transitions (blue solid). (c) and (f) show the populations in the
relevant dressed states for varied driving powers $\Omega_{d}^{2}/\gamma_{10}^{2}$.
To help identify where the amplification without population inversion
takes place, we plot the associated (maximum) reflection amplitude
for reference.\textcolor{red}{\label{fig: double_sideband}}}
\end{figure*}

Now we examine the effect of multiple pump-photon resonance, amounting
to setting $\omega_{d}=\left(\omega_{K}-\omega_{0}\right)/K$, and
take $K=2$ and $K=3$ cases as a workhorse towards understanding
of the gain effects. Figure~\ref{fig: double_sideband}(a) displays
the reflection amplitude for $K=2$. A resonant curve observed around
$\omega_{p}=\omega_{10}$ for weak pumping shows attenuation until
it splits into two branches with gain $r>1$ for $\Omega_{d}^{2}/\gamma_{10}^{2}\gtrsim10^{2}$.
The splitting signals are the formation of dressed states due to a
mechanism similar to that observed in the single pump-photon resonant
case. Meanwhile, we also notice that a spectral curve emerges around
$\omega_{p}-\omega_{10}=-\alpha/2=-50\gamma_{10}$, corresponding
to the two-probe-photon resonance with the transition $|2\rangle\leftrightarrow|0\rangle$,
and splits into a branch of amplification and the other one of attenuation.
In these cases, population inversion is the only factor responsible
for the amplification of the probe signal.

On the other hand, another curve merging from $\omega_{p}-\omega_{10}=-\alpha=-100\gamma_{10}$
splits into two branches of attenuation as $\Omega_{d}^{2}/\gamma_{10}^{2}$
exceeds $2\times10^{1}$. Interestingly, it is noteworthy that when
$\Omega_{d}^{2}/\gamma_{10}^{2}\approx10^{2}$, we find up to a $10$\%
signal amplification result from the interference of these two attenuation
curves. Our analysis shows that such amplification is actually owing
to the interference of the dressed-state transitions $\left|D_{3},F+1\right\rangle \leftrightarrow\left|D_{5},F\right\rangle $
and $\left|D_{4},F+1\right\rangle \leftrightarrow\left|D_{5},F\right\rangle $.
This can be supported by Eq.~(\ref{eq: double_sidebands_EQ}), dealing
with the mixture of two spectral contributions: We manually cut out
the coupling, amounting to setting to zero the off-diagonal terms
of the coefficient matrix on the LHS of Eq.~(\ref{eq: double_sidebands_EQ}),
and find that the amplification is removed with $r\rightarrow1$.
See Fig.~\ref{fig: double_sideband}(b). Further, we look into the
populations of the relevant states along with the reflection. Figure~\ref{fig: double_sideband}(c)
shows the population distributions, which are dominantly determined
by the pump intensity, and the maximum gain observed between the two
attenuation curves. One can clearly spot the gain without the population
inversion for $\Omega_{d}^{2}/\gamma_{10}^{2}\gtrsim10^{2}$, suggesting
that the amplification is mostly due to the interference of the nearly
degenerate transitions.

We want to highlight a scenario in which, even in the presence of
population inversion, interference can additionally contribute to
enhanced gain. This occurs when two amplifying transitions intersect.
A notable example is illustrated in Fig.~\ref{fig: double_sideband}(d)
for $K=3$, where $\Omega_{d}^{2}/\gamma_{10}^{2}=5.6\times10^{3}$.
Here, the transitions $\left|D_{5},F+1\right\rangle \leftrightarrow\left|D_{4},F\right\rangle $
and $\left|D_{4},F+1\right\rangle \leftrightarrow\left|D_{3},F\right\rangle $
become degenerate. If we disregard the coupling among near-resonant
Rabi sidebands, the anticipated overall gain is $12.5$\%, calculated
as the direct sum of individual transition contributions. However,
the maximum achievable gain from a complete calculation can reach
up to $21.5\%$. This enhancement, reported in a recent experiment
\citep{Aziz2023}, aligns with our theory and presents a finding not
discussed in previous literature.

\subsection{Effect of the pure dephasing process \label{subsec:Effect-of-the}}

\begin{figure}
\begin{centering}
\includegraphics[width=6cm]{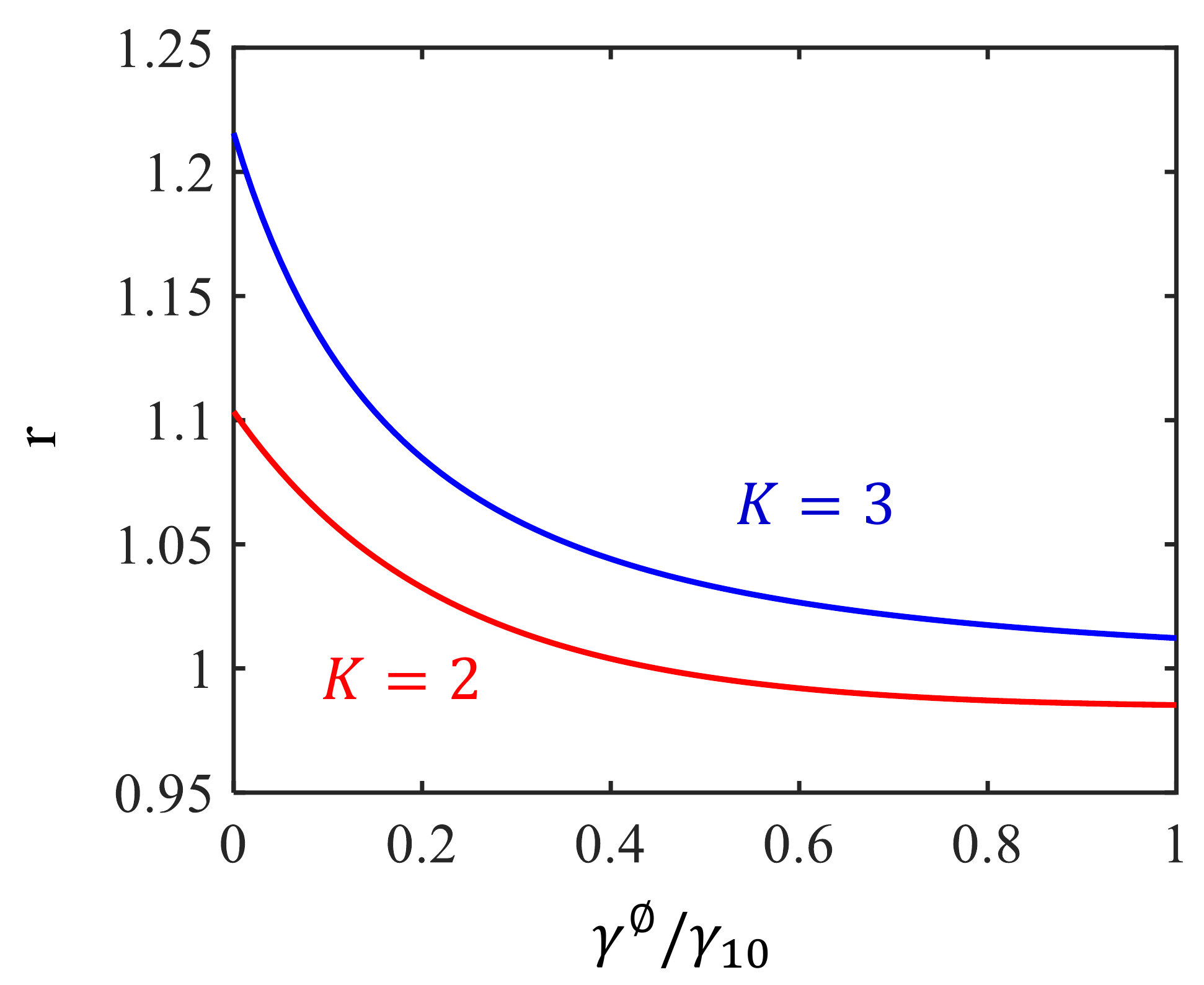}
\par\end{centering}
\caption{Reflection amplitudes for different dephasing rates for both the $K=2$
and $K=3$ cases. We use $\Omega_{d}^{2}/\gamma_{10}^{2}=10^{2}$
and $\omega_{p}=\omega_{10}-102\gamma_{10}$ for $K=2$, and $\Omega_{d}^{2}/\gamma_{10}^{2}\approx5.6\times10^{3}$
and $\omega_{p}=\omega_{10}-12.1\gamma_{10}$ for $K=3$. \label{fig: pure_dephasing_process}}
\end{figure}

Finally, we incorporate the dephasing effect into our calculations
to better represent realistic setups. Figure~\ref{fig: pure_dephasing_process}
displays the cases for $K=2$ and $K=3$ with various pure dephasing
rates. Generally, we observe that the maximum amplitude occurs at
zero dephasing, and the reflection amplitude decreases as the dephasing
rate increases. Employing a pure dephasing rate of $\gamma^{\phi}/\gamma_{10}\approx0.014$,
as measured in \citep{Aziz2023}, we obtain $r\approx1.18$. This
result precisely aligns with the experimental data, thereby validating
our theoretical model.

\section{Conclusion\label{sec: Conclusion}}

In conclusion, our study focused on the gain mechanism in a signal
amplifier consisting of a multi-level transmon driven by a strong
coherent field. This intense field interacts with the higher energy
levels of the transmon, resulting in the formation of multiple dressed
states, or Rabi sidebands. We observed significant amplification of
a weak probe signal through its interaction with these sidebands.
To elucidate this gain mechanism, we developed a comprehensive theoretical
model that incorporates the dressed-state representation and accounts
for cross-sideband coherences. Our findings indicate that amplification
can arise from either population inversion or stimulated photon constructive
interference involving multiple sidebands. For example, in the two
pump-photon resonant case, we achieved amplification without population
inversion. In the three-photon scenario, we found that the gain could
be further enhanced by interference among multiple sidebands, in addition
to the effects of population inversion. Notably, our calculations
align excellently with recent experimental observations by \citep{Aziz2023},
validating our theoretical approach. We also explored the impact of
pure dephasing on amplification. Overall, our research provides valuable
insights into amplification mechanisms at the single-atom level and
lays a solid foundation for future development in quantum repeaters,
interfaces, and networks, thereby advancing the field of quantum communications.

\section*{Appendix}

\subsection{The multi-transmon master equation and reflection amplitude}

\setcounter{equation}{0} \renewcommand{\theequation}{A.\arabic{equation}}

In this section, we extend our model from a single transmon to a multi-transmon
system, which consists of a linear array of $N$ identical transmons
coupled to a one-dimensional (1D) semi-infinite waveguide. As outlined
in Sec.~\ref{sec: Theory}, we assume each transmon has $J$ energy
levels. An intense coherent driving field is applied from the open
end, exciting each transmon from the ground state $\left|0\right\rangle $,
to a state $\left|K\right\rangle $ via a $K$ pump-photon process.
Similar to the discussion in Sec.~\ref{sec: Theory}, we start with
the Hamiltonian $H=H_{s}+H_{b}+V+H_{d}+H_{p}$, where the atomic part
becomes $H_{s}=\sum_{n,j}\hbar\omega_{j}\sigma_{jj}^{n}$ dealing
with multiple transmons for $n=1,2,\cdots,N$, and their interaction
\begin{equation}
V=i\hbar\sum_{n,j}\int_{0}^{\infty}d\omega\sqrt{j}g\left(\omega\right)\cos\left(k_{\omega}x_{n}\right)\sigma_{j,j-1}^{n}a_{\omega}+{\rm H.c.}\label{eq: multi-atom V}
\end{equation}
assuming the $n$th transmon is located at position $x=x_{n}$ with
its associated operator $\sigma_{j,k}^{n}=\left|j\right\rangle _{n}\left\langle k\right|$
for $j,k=1,2,\ldots J$. Further, the coupling of the driving and
probe fields to this array of transmons is represented, respectively,
by 
\begin{equation}
H_{d}=\hbar\sum_{n,j}\frac{\sqrt{j}\Omega_{d}\cos\left(k_{d}x_{n}\right)}{2}\left(\sigma_{j,j-1}^{n}e^{-i\omega_{d}t}+{\rm H.c.}\right)\label{eq: multi-aton Hd}
\end{equation}
and
\begin{equation}
H_{p}=\hbar\sum_{n,j}\frac{\sqrt{j}\Omega_{p}\cos\left(k_{p}x_{n}\right)}{2}\left(\sigma_{j,j-1}^{n}e^{-i\omega_{p}t}+{\rm H.c.}\right).\label{eq: multi-atom pump}
\end{equation}
By eliminating the waveguide photonic modes, we derive the Born-Markov
master equation in the rotating frame of frequencies $\omega_{d}$:
\begin{equation}
\begin{aligned}\frac{d\rho}{dt} & =i\sum_{n,j}\left(j\omega_{d}-\omega_{j}\right)\left[\sigma_{jj}^{n},\rho\right]\\
 & +i\sum_{n,j}\frac{\sqrt{j}\Omega_{d}\cos\left(k_{d}x_{n}\right)}{2}\left(\left[\sigma_{j,j-1}^{n},\rho\right]-{\rm H.c.}\right)\\
 & +i\sum_{n,j}\frac{\sqrt{j}\Omega_{p}\cos\left(k_{p}x_{n}\right)}{2}\left(\left[\sigma_{j,j-1}^{n},\rho\right]e^{i\left(\omega_{d}-\omega_{p}\right)t}-{\rm H.c.}\right)\\
 & +i\sum_{nm,jl}\sqrt{jl}\Delta_{j,j-1}^{nm}\left(\left[\sigma_{j-1,j}^{m}\rho,\sigma_{l,l-1}^{n}\right]-{\rm H.c.}\right)\\
 & +\sum_{nm,jl}\frac{\sqrt{jl}\gamma_{j,j-1}^{nm}}{2}\left(\left[\sigma_{j-1,j}^{m}\rho,\sigma_{l,l-1}^{n}\right]+{\rm H.c.}\right)\\
 & +\sum_{n,j}\gamma_{n,j}^{\phi}\left(\left[\sigma_{jj}^{n}\rho,\sigma_{jj}^{n}\right]+{\rm H.c.}\right).
\end{aligned}
.\label{eq: multi-atom master equatuin}
\end{equation}
Here, the collective decay rate and Lamb shift \citep{Lehmberg1970,Gross1982,Lin2019},
accounting for the dipole-dipole interaction between the $n$th and
$m$th transmons, are given by
\begin{equation}
\gamma_{j,j-1}^{nm}=\frac{\gamma_{j,j-1}}{2}\text{{\rm Re}}\left(e^{ik_{j,j-1}\left(x_{n}+x_{m}\right)}+e^{ik_{j,j-1}\left|x_{n}-x_{m}\right|}\right)\label{eq: collective decay rates}
\end{equation}

\begin{equation}
\Delta_{j,j-1}^{nm}=\frac{\gamma_{j,j-1}}{4}\text{{\rm Im}}\left(e^{ik_{j,j-1}\left(x_{n}+x_{m}\right)}+e^{ik_{j,j-1}\left|x_{n}-x_{m}\right|}\right)\label{eq: collective Lamb shift}
\end{equation}
with the same notations specified in the main text.

Again, the input and output signal can be related through the expression:

\begin{equation}
a_{{\rm out}}\left(t\right)=a_{{\rm in}}\left(t\right)+\sum_{n,j}\sqrt{j\gamma_{j,j-1}}\cos\left(k_{j,j-1}x_{n}\right)\sigma_{j-1,j}^{n}\left(t\right),\label{eq: multi-atom output}
\end{equation}
and the reflection amplitude is given by
\begin{equation}
r=\left|1+\frac{2i\sum_{n,j}\tilde{\gamma}_{j}\left(x_{n}\right)\left\langle \tilde{\sigma}_{j-1,j}^{n}\left(t\right)\right\rangle e^{i\left(\omega_{p}-\omega_{d}\right)t}}{\Omega_{p}}\right|,\label{eq: r_multi-atom}
\end{equation}
where $\tilde{\sigma}_{j-1,j}^{n}\left(t\right)=\sigma_{j-1,j}^{n}\left(t\right)e^{i\omega_{d}t}$
denotes the atomic ladder operator for $n$th transmon in the rotating
frame of frequency $\omega_{d}$. The expectation value is then determined
via $\left\langle \tilde{\sigma}_{j-1,j}^{n}\left(t\right)\right\rangle ={\rm Tr}\left(\Omega_{p}\tilde{\sigma}_{j-1,j}^{n}\rho^{\left(1\right)}\left(t\right)e^{-i\left(\omega_{p}-\omega_{d}\right)t}/\gamma_{10}\right)$.

\subsection{Multi-atom dressed state approach}

\setcounter{equation}{0} \renewcommand{\theequation}{B.\arabic{equation}}

This section builds up the dressed-state representation similarly
in Sec.~\ref{sec: Theory}. We reformulate the reflection amplitude
as
\begin{equation}
r=\left|1+2i\sum_{\mu\nu}C_{D_{\mu}D_{\nu}}\left\langle \sigma_{D_{\mu}D_{\nu}}\left(t\right)\right\rangle \right|.\label{eq: multi-atom dressed-state r}
\end{equation}
Here, the coefficient $C_{D_{\mu}D_{\nu}}$ is determined by the ladder
operator through $C_{D_{\mu}D_{\nu}}=\sum_{n,j}\tilde{\gamma}_{j}\left(x_{n}\right){\rm Tr}\left(\sigma_{j-1,j}^{n}\sigma_{D_{\nu}D_{\mu}}\right)/\gamma_{10}$,
where $\sigma_{D_{\mu}D_{\nu}}=\left|D_{\nu}\right\rangle \left\langle D_{\mu}\right|=\sigma_{D_{\mu}D_{\nu}}^{\dagger}$
and its dynamics are governed by the equation: \begin{widetext}
\begin{equation}
\begin{aligned}\frac{d\left\langle \sigma_{D_{\mu}D_{\nu}}\right\rangle }{dt} & =\left(i\delta_{\mu\nu}^{D}+\Gamma_{\mu\nu}^{D}\right)\left\langle \sigma_{D_{\mu}D_{\nu}}\right\rangle \\
 & +i\sum_{\eta}\left(\Omega_{\eta\nu}^{D}\left\langle \sigma_{D_{\mu}D_{\eta}}\right\rangle _{\left(0\right)}-\Omega_{\mu\eta}^{D}\left\langle \sigma_{D_{\eta}D_{\nu}}\right\rangle _{\left(0\right)}\right)\\
 & +\sum_{\xi\neq\mu,\eta\neq\nu}\bar{\Gamma}_{\eta\nu\mu\xi}^{D}\left\langle \sigma_{D_{\xi}D_{\eta}}\right\rangle -\sum_{\eta,\xi\neq\nu}\Gamma_{\eta\nu\xi\eta}^{D^{+}}\left\langle \sigma_{D_{\mu}D_{\xi}}\right\rangle -\sum_{\xi,\eta\neq\mu}\Gamma_{\xi\eta\mu\xi}^{D^{-}}\left\langle \sigma_{D_{\eta}D_{\nu}}\right\rangle 
\end{aligned}
.\label{eq: multi-atom OBE}
\end{equation}
\end{widetext} Here, the term $\delta_{\mu\nu}^{D}=\omega_{p}-\left(\omega_{\nu}^{D}-\omega_{\mu}^{D}+\omega_{d}\right)$
denotes the detuning and $\Gamma_{\mu\nu}^{D}=\bar{\Gamma}_{\nu\nu\mu\mu}^{D}-\sum_{\eta}\left(\Gamma_{\eta\nu\nu\eta}^{D^{+}}+\Gamma_{\eta\mu\mu\eta}^{D^{-}}\right)$
represents relaxation rate, including contributions from 
\begin{equation}
\begin{aligned}\bar{\Gamma}_{\eta\nu\mu\xi}^{D} & =\sum_{nm,jk}\chi_{jk}^{nm}{\rm Tr}\left[\sigma_{k-1,k}^{m}\sigma_{D_{\eta}D_{\nu}}\right]{\rm Tr}\left[\sigma_{j,j-1}^{n}\sigma_{D_{\mu}D_{\xi}}\right]\\
 & \qquad+2\sum_{n,j}\gamma_{n,j}^{\phi}{\rm Tr}\left[\sigma_{jj}^{n}\sigma_{D_{\eta}D_{\nu}}\right]{\rm Tr}\left[\sigma_{jj}^{n}\sigma_{D_{\mu}D_{\xi}}\right]
\end{aligned}
\label{eq: multi-atom Gamma_bar}
\end{equation}
and
\begin{equation}
\begin{aligned}\Gamma_{\eta\nu\mu\xi}^{D^{\pm}} & =\sum_{nm,jk}\eta_{nm,jk}^{\pm}{\rm Tr}\left[\sigma_{j,j-1}^{n}\sigma_{D_{\eta}D_{b}}\right]{\rm Tr}\left[\sigma_{k-1,k}^{m}\sigma_{D_{\mu}D_{\xi}}\right]\\
 & +\sum_{n,j}\gamma_{n,j}^{\phi}{\rm Tr}\left[\sigma_{jj}^{n}\sigma_{D_{\eta}D_{\nu}}\right]{\rm Tr}\left[\sigma_{jj}^{n}\sigma_{D_{\mu}D_{\xi}}\right]
\end{aligned}
,\label{eq: multi-atom Gamma_pn}
\end{equation}
where $\chi_{jk}^{nm}=\sqrt{jk}\left(\frac{\gamma_{k,k-1}^{nm}+\gamma_{j,j-1}^{nm}}{2}-i\left(\Delta_{k,k-1}^{nm}-\Delta_{j,j-1}^{nm}\right)\right)$
and $\eta_{nm,jk}^{\pm}=\sqrt{jk}\left(\frac{\gamma_{k,k-1}^{nm}}{2}\pm i\Delta_{k,k-1}^{nm}\right)$.
The optical pumping process is further characterized by $\Omega_{\mu\nu}^{D}\left\langle \sigma_{D_{\mu}D_{\nu}}\right\rangle _{\left(0\right)}=\sum_{n,j}\frac{\sqrt{j}\gamma_{10}}{2}\cos\left(k_{p}x_{n}\right){\rm Tr}\left[\sigma_{j,j-1}^{n}\sigma_{D_{\mu}D_{\nu}}\right]{\rm Tr}\left[\rho^{\left(0\right)}\left|D_{\nu}\right\rangle \left\langle D_{\mu}\right|\right]$
while the third row of Eq.~(\ref{eq: multi-atom OBE}) determines
the coupling between the dressed states. Each dressed state, $\left|D_{\mu}\right\rangle $,
is an eigenstate of the interaction-picture multi-atom Hamiltonian
$H_{s}+H_{d}$, with $\mu=0,1,2\cdots,\left(J-1\right)^{N}$.

Anticipating significant contributions from numerous near-resonant
sidebands on the reflection signal, we explore conditions under which
these sidebands, characterized by $\delta_{\mu\nu}^{D}\lesssim\gamma_{10}$
for $\mu,\nu=0,1,2\ldots,\left(J-1\right)^{N}$ with $\mu\neq\nu$,
lead to interference effects on the optical response of the probe
field. For illustration, consider a scenario where $M$ dressed-state
transitions, denoted by $\left|D_{\nu_{i}},F+1\right\rangle \leftrightarrow\left|D_{\mu_{i}},F\right\rangle $
with $i=0,1,2\ldots,M\leq\left(J-1\right)^{N},$ are resonant to the
probe field. From Eq.~(\ref{eq: multi-atom OBE}), we derive a set
of coupled equations for the steady state:\begin{widetext}
\begin{equation}
\Pi\left[\begin{array}{c}
\left\langle \sigma_{D_{\mu_{1}}D_{\nu_{1}}}\right\rangle \\
\left\langle \sigma_{D_{\mu_{2}}D_{\nu_{2}}}\right\rangle \\
\vdots\\
\left\langle \sigma_{D_{\mu_{M}}D_{\nu_{M}}}\right\rangle 
\end{array}\right]=\left[\begin{array}{c}
i\sum_{\eta}\left(\Omega_{\mu_{1}\eta}^{D}\left\langle \sigma_{D_{\eta}D_{\nu_{1}}}\right\rangle _{\left(0\right)}-\Omega_{\eta\nu_{1}}^{D}\left\langle \sigma_{D_{\mu_{1}}D_{\eta}}\right\rangle _{\left(0\right)}\right)\\
i\sum_{\eta}\left(\Omega_{\mu_{2}\eta}^{D}\left\langle \sigma_{D_{\eta}D_{\nu_{2}}}\right\rangle _{\left(0\right)}-\Omega_{\eta\nu_{2}}^{D}\left\langle \sigma_{D_{\mu_{2}}D_{\eta}}\right\rangle _{\left(0\right)}\right)\\
\vdots\\
i\sum_{\eta}\left(\Omega_{\mu_{M}\eta}^{D}\left\langle \sigma_{D_{\eta}D_{\nu_{M}}}\right\rangle _{\left(0\right)}-\Omega_{\eta\nu_{M}}^{D}\left\langle \sigma_{D_{\mu_{M}}D_{\eta}}\right\rangle _{\left(0\right)}\right)
\end{array}\right].\label{eq: M sideband EQ}
\end{equation}
\end{widetext}Here, $\Pi$ is the coefficient matrix, with each element
$\Pi_{ij}=\left(\Gamma_{\mu_{i}\nu_{j}}^{D}+i\delta_{\mu_{i}\nu_{j}}^{D}\right)\delta_{ij}+\Gamma_{\nu_{j}\nu_{i}\mu_{i}\mu_{j}}^{D}$,
highlighting the interference between dressed-state transitions as
denoted by the non-diagonal terms. The symbol $\delta_{ij}$ represents
the Kronecker delta function.
\begin{acknowledgments}
K.-T. L. acknowledges Chen-Yu Lee and Chun-Chi Wu for the helpful
discussions and the support from the National Science and Technology
Council of Taiwan under Grant No. 112-2811-M-002-067. G.-D. L. acknowledges
the support from the National Science and Technology Council of Taiwan
under Grant No. 112-2112-M-002-001. P.-Y. W. acknowledges the support
from the National Science and Technology Council of Taiwan under Grant
No. 110-2112-M-194-006-MY3.
\end{acknowledgments}

\bibliography{amplification_citation}

\end{document}